\documentclass[prl,aps,amsmath,twocolumn,unsortedaddress,altaffilletter,superscriptaddress]{revtex4}
\usepackage{natbib}
\usepackage{graphicx}
\usepackage{multirow}
\usepackage{verbatim}
\usepackage{physics}
\usepackage{textcomp}

\usepackage{cancel}
\usepackage[normalem]{ulem}
\usepackage{mathtools}
\usepackage{amsfonts}

\usepackage{siunitx}
\usepackage{MnSymbol}

\usepackage{color}

\usepackage{rotating}

\usepackage{amsmath}

\usepackage{wrapfig}

\usepackage[plainpages=false]{hyperref}

\begin{document}

\title{Pinning of Diffusional Patterns by Non-Uniform Curvature}
\date{\today}

\author{John R. Frank}
\email{jrf@mit.edu}
\affiliation{Dept. of Physics, Massachusetts Institute of Technology, Cambridge, MA 02138, USA}

\author{Jemal Guven}
\affiliation{Instituto de Ciencias Nucleares, Universidad Nacional Aut\'{o}noma de M\'{e}xico,
Apdo. Postal 70-543, 04510 M\'{e}xico, DF, MEXICO}

\author{Mehran Kardar}
\affiliation{Dept. of Physics, Massachusetts Institute of Technology, Cambridge, MA 02138, USA}

\author{Henry Shackleton}
\email{hshackleton@g.harvard.edu}
\affiliation{Dept. of Physics, Harvard University, Cambridge, MA 02138, USA}

\begin{abstract}
Diffusion-driven patterns appear on curved surfaces in many settings, 
initiated  by unstable modes of an underlying Laplacian operator.
On a  flat surface or perfect sphere, the patterns are degenerate, reflecting 
translational/rotational symmetry. 
Deformations, e.g. by a bulge or indentation, break symmetry and can pin a pattern. 
We adapt methods of conformal mapping and perturbation theory to examine how 
curvature inhomogeneities select and pin patterns, and confirm the results numerically. 
The theory provides an analogy to quantum mechanics in a geometry-dependent potential
and yields intuitive implications for cell membranes, tissues, thin films, and noise-induced quasipatterns.


\end{abstract}


\maketitle

In 1952, Turing coined the term ``morphogen'' in a seminal paper that
showed how combining diffusion with generalized reactions can create spatial and temporal patterns
even though separately each leads to uniform, static concentrations~\cite{Turing:1952p2540}. 
Such reaction-diffusion (RD) patterns are but one example of diffusion-driven instabilities that have since
been studied on scales ranging from
microns in cells~\cite{hecht2010transient}
and active fluids~\cite{bois2011pattern, giomi2012polar, pearce2018geometrical, durre2018capping, marchetti2013hydrodynamics, kumar2014pulsatory},
millimeters in hydrodynamics~\cite{almarcha2010chemically},
centimeters in zoology~\cite{Murray:1981b}, to meters in ecology~\cite{Butler:2011gh}.
Diffusion-driven patterns are known to determine morphology~\cite{kondo2010reaction} 
in model organisms like zebrafish~\cite{nakamasu2009interactions, yamanaka2014vitro} 
and complex organs like the eye~\cite{sasai2013cytosystems}.
Recent theoretical progress in patterning~\cite{Biancalani:2017, Corson:2017}
encourages further study.

Substrate curvature plays a role in pattern formation in many systems, 
including cell membranes~\cite{isaac2013linking} and 
thin films~\cite{turner2010vortices}.
The importance of surface curvature on collective behavior has recently been explored in liquid crystals~\cite{Fialho:2017}, flocking~\cite{Shankar:2017}, and wave propagation~\cite{PhysRevLett.120.268001}. 
Closer to our work, the geometric dependence of pattern formation has recently been studied in various models of protein~\cite{Thalmeier:2016, Vandin:2016} and molecular bonding~\cite{Yu:2017}. 

Recent studies of Turing patterns have explored the effects of curvature on highly symmetric shapes such as spheres, cylinders, toroids~\cite{Nampoothiri:2016,Sanchez-Garduno:2018}, and ellipsoids~\cite{Nampoothiri:2017} where the Laplacian is known in closed form. Inhomogeneities in curvature, such as protrusions or cavities, reduce such symmetries
and can pin or modify the patterns.
To understand how nonuniform curvature can entrain and modify patterns, we study perturbations to the Laplacian,
and its eigenmodes.
To our knowledge, the intimate link between pinning of patterns and the spectrum of the Laplacian has not been pursued. 

We follow a two prong strategy. 
First, we identify the {\it onset} of instabilities, by linearizing evolution equations
expressed in terms of the appropriate `modes', e.g., Fourier, cylindrical, or spherical harmonics.
Modes with the largest positive real part grow fastest and are harbingers of the final patterns molded by nonlinearities.
The modes are eigenfunctions of the diffusion (Laplacian) operator on the relevant manifold.
Symmetries of the manifold, reflected in degeneracies of the eigenfunctions, must be broken
in the final patterns.
Previous work on the Laplacian on Riemannian manifolds focused
on its determinant~\cite{Kac:1966p4109,Schoen:1994,Rosenberg:1997,Braun:2008p4810} 
and short-time behavior appropriate to field theory~\cite{David:1988,Faraudo:2002p5195,CastroVillarreal:2010p6177}.
We focus instead on how non-uniform curvature breaks degeneracies,
pinning eigenfunctions to inhomogeneities. 
To do this, we utilize conformal mappings and perturbation theory.
There is no guarantee, however, that patterns resulting from non-linear evolution
are similarly entrained, so the second step in our study explores patterns with simulations. 
We implemented the Thomas-Murray RD equations~\cite{Murray:1981b} on COMSOL Multiphysics\textregistered~\cite{comsol}.
We conclude with suggestions for experiments.

We first consider a cylinder
with axially symmetric deformations described in cylindrical polar coordinates as $\rho = R(z)$. 
The surface line element is 
\begin{equation}\label{eq:CylLine1}
  \dd{s}^2 = (1+R_z^2) \dd{z}^2 + R^2 \dd{\varphi}^2\,,
\end{equation}
where $R_z$ denotes the derivative with respect to $z$. 
The Laplacian of a scalar $\phi$ is
\begin{eqnarray}
  \Delta \phi
&=& \frac{1}{\sqrt{g}} \partial_a \left( \sqrt{g} g^{ab}  \partial_b \phi \right)
\label{Conformal:LaplacianByMetric} \text{  ,}
\end{eqnarray}
where $g_{ab}$ is the metric of the underlying geometry; $g$ is the determinant of the metric. 
Conformal mapping simplifies analysis through mapping to a flat geometry.
We introduce a conformal axial coordinate $v$, such that the line element 
acquires the conformally flat form
\begin{equation}\label{eq:CylLine2}
  \dd{s}^2 = \Omega^2(v) \left(\dd{v}^2 + R_0^2 \dd{\varphi}^2\right)\,,
\end{equation}
where $R_0$ is the asymptotic radius and $\Omega$ is the conformal factor. 
In the conformal coordinates, $\sqrt{g} = \Omega^2 R_0$, and thus the Laplacian on the deformed geometry takes the
simple form $\Delta^G = \Omega^{-2} \Delta^0$, where $\Delta^0$ is the Laplacian in the conformally flat coordinates.
Since the behavior of $\Delta^0$ is well understood, and $\Omega$ is determined by the equality of Eqs.~\ref{eq:CylLine1} and \ref{eq:CylLine2}, this conformal mapping provides a tractable method of understanding $\Delta^G$. 

Solutions for $v$ and $\Omega$ for arbitrary $R(z)$ are in the supplement.
To develop perturbation theory, we set $R = R_0(1+\epsilon h(z))$, in which case
$v\approx z$ and $\Omega\approx 1+2 \epsilon h(v)$ to lowest order, such that
the eigenfunctions  $\phi_k$ of $\Delta^G$, with eigenvalues $\lambda$,  satisfy 
\begin{equation}
  -\Delta^0 \phi_k + 2 \lambda  \epsilon h(v) \phi_k = -\lambda \phi_k\,.
\label{geneigenvalue}
\end{equation}
In analogy with quantum mechanics, one can interpret the deformation as giving rise to a potential in the conformal coordinates, whose magnitude is dependent on the eigenvalue $\lambda$.
(This differs from da Costa's geometric potential, which comes from confining a particle to a surface \cite{daCosta:1982}.)
For physical phenomena described by the Laplacian, this mapping can be interpreted as the replacement 
of the diffusion operator $-D \Delta$ on a deformed geometry,  with a spatially-dependent diffusion coefficient
$\tilde{D}(v) \equiv D/\Omega^{2}$ in the conformally-related homogeneous geometry.
This provides an intuitive picture of how diffusion is modified by curvature.

The undistorted cylinder of length $L$ with periodic boundary conditions
has eigenfunctions and eigenvalues
\begin{equation}
  \begin{aligned}
    \phi^{(0)}_{sk} &= \frac{e^{is\varphi}}{\sqrt{2\pi R_0}} \frac{e^{i 2\pi k z/L}}{\sqrt{L}}\,,
    \\
    \lambda^{(0)}_{sk} &= -\frac{s^2}{R_0^2} - \left(\frac{2\pi k}{L}\right)^2 \equiv -\bar{s}^2 - \bar{k}^2\,,
	\nonumber
  \end{aligned}
\end{equation}
with integers $s$ and $k$. 
Consider an axially-symmetric bump on the cylinder, shaped like a Gaussian of standard deviation $\sigma$ and height $v_0$. 
We apply Rayleigh-Schr\"{o}dinger perturbation theory to Eq.~\ref{geneigenvalue} to calculate eigenvalue corrections
to first order in $\epsilon \equiv v_0/R$, which simplify for the case $s=0$ to
\begin{equation}
  \begin{aligned}
    \lambda_{0k}^{\pm}&= -\bar{k}^2 \left(1 \pm \frac{ 2\epsilon \sigma \sqrt{2\pi}}{L} e^{-2 \bar{k}^2 \sigma^2} \right) + \order{\epsilon^2}\,.
\end{aligned}
\nonumber
\end{equation}
The positive/negative sign is set by the mode: positive for cosine (symmetric) modes, 
and negative for sine (antisymmetric) modes. 
The sign of $\epsilon$ depends on the orientation of the ridge-like deformation:
positive for a bulge and negative for a constriction. 
Thus, to first order, a ridge breaks the degeneracy in the eigenvalues of a Laplacian on the cylinder, 
causing the eigenvalues of sine (cosine) modes to become more negative in the case of an outward (inward) ridge. 
This correction is largest for $2\pi/\bar{k} = 4 \pi \sigma$, when the wavelength is approximately twice the width of the ridge. 
This is a general trait of deformations, explored later for a rippled cylinder. 
The ridge can be positioned anywhere along the cylinder, and the modes will shift with the ridge. 

Modifications to the Laplacian spectrum affect any physical system involving diffusion.  
For example, consider a two-component RD system:
\begin{eqnarray}
	\partial_ t
	\left(\begin{array}{c} \Psi_1({\bf x}, t) \\ \Psi_2({\bf x}, t)\end{array}\right) 
	=
	\overbrace{
	\left(\begin{array}{c} 
		R_1\left(\Psi_1, \Psi_2 \right) \\ R_2\left(\Psi_1, \Psi_2 \right)
	\end{array}\right) 
	}^{\text{Reactions}}
	+
	\overbrace{
	\left(\begin{array}{cc} 
		\nu_1 & 0 \\ 0 & \nu_2
	\end{array}\right) 
	\Delta^G
	\left(\begin{array}{c} \Psi_1 \\ \Psi_2\end{array}\right) 
	}^{\text{Diffusion}}\,.
	\label{eqn:RD}
\end{eqnarray}
After linearization, small deviations around a uniform stable fixed point of the reactions, $(\Psi_1,\Psi_2)^*$, evolve as
\begin{eqnarray}
	 \left(\begin{array}{c} \Psi_1({\bf x}, t) \\ \Psi_2({\bf x}, t)\end{array}\right) 
	= \left(\begin{array}{c} \Psi_1 \\ \Psi_2\end{array}\right)^*\! + \displaystyle \sum_k \left(\begin{array}{c} u_{1k} \\ u_{2k} \end{array}\right) e^{t \omega(k)}\phi_k\left({\bf x}\right) 
	\nonumber
	\text{  ,}
\end{eqnarray}
with $\omega(k)$ satisfying eigenvalue equation, $\vb{M}(\lambda) = \vb{R} + \lambda \boldsymbol\nu$,
\begin{eqnarray}
 \left[
\left(\begin{array}{cc} R_{1,1} & R_{1,2} \\ R_{2,1} & R_{2,2}\end{array}\right)^* \!\!
+
\left( \begin{array}{cc} \nu_1 \lambda_k & 0 \\ 
0 & \nu_2 \lambda_k \end{array} \right)  
\right]
\left(\begin{array}{c} u_{1k} \\ u_{2k} \end{array}\right)
=\omega(k)\left(\begin{array}{c} u_{1k} \\ u_{2k} \end{array}\right)
\nonumber
\text{.}
\end{eqnarray}
While the diffusion and stable reaction matrices separately posses negative
eigenvalues, Turing showed that their sum can have positive eigenvalues
 ($\omega_+(k)>0$) signaling finite-wavelength instabilities~\cite{Turing:1952p2540}.  
The possibly degenerate modes, $k^*$, with largest eigenvalue evolve to the final pattern. 
On a cylinder with a ridge, the degeneracy between
sine and cosine modes is broken: 
an outward (inward) ridge leads to sine (cosine) growing faster.

\begin{figure}[h]
  \includegraphics[width=0.4\textwidth, height=0.27777\textwidth]{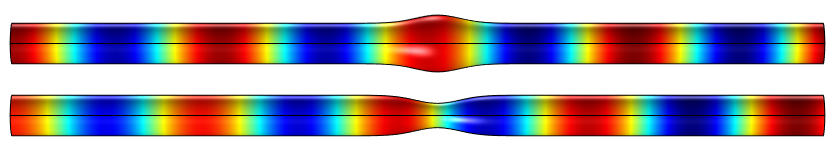}
  \includegraphics[width=0.05\textwidth]{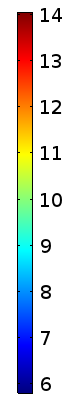}
  \caption{Patterns from the Thomas-Murray RD model in Eq.~\eqref{eq:Thomas-Murray}
are entrained to a Gaussian-shaped ridge, switching in phase between inward and outward deformations. Red (blue) indicates a high (low) concentration of chemicals. Vertical dimension magnified 3x.
Parameters:
$\nu_1 = 1$, $\nu_2 = 10$, $u_{10} = 92$, $u_{20} = 64$, $\alpha = 1.5$, $K=0.1$, $\gamma = 2$, and $\rho = 18.5$.
Unless specified otherwise, other figures have these parameters.}
  \label{fig:cylinder-bump}
\end{figure}

This linear analysis indicates only the onset of instabilities.  To
understand the patterns formed after nonlinearities stabilize the dynamics, we conducted finite
element simulations of the Thomas-Murray model~\cite{Murray:1981b}:
\begin{eqnarray}
	\dot{\Psi}_1 &=& \nu_1 \Delta^G \Psi_1 + \gamma\left( \Psi_1 - u_{10} - \frac{\rho \Psi_1 \Psi_2}{1 + \Psi_1 +  \Psi_1^2/K}\right)\,
	\nonumber \\
	\dot{\Psi}_2 &=& \nu_2 \Delta^G \Psi_2 + \gamma\left( \alpha\left(\Psi_2 - u_{20}\right) - \frac{\rho \Psi_1 \Psi_2}{1 + \Psi_1 +  \Psi_1^2/K} \right)
	\label{eq:Thomas-Murray} \text{  .}
\end{eqnarray}
We used COMSOL Multiphysics\textregistered \cite{comsol}, which approximates the Laplacian by
finite differences on a mesh and computes the fully non-linear
reaction terms.  (Supplement includes .mph file.)
It should be noted that, although periodic boundary conditions are enforced in the numerical
calculations, the computational methods using by COMSOL have a tendency to pin the resulting Turing
patterns in a specific configuartion, even on a flat cylinder.
Despite this, as shown in Fig.~\ref{fig:cylinder-bump}, density maxima of Turing patterns on a deformed cylinder  
are entrained by a ridge.  
An inward (outward) ridge selects the sine (cosine) mode.

\begin{figure}
  \includegraphics[width=0.5\textwidth]{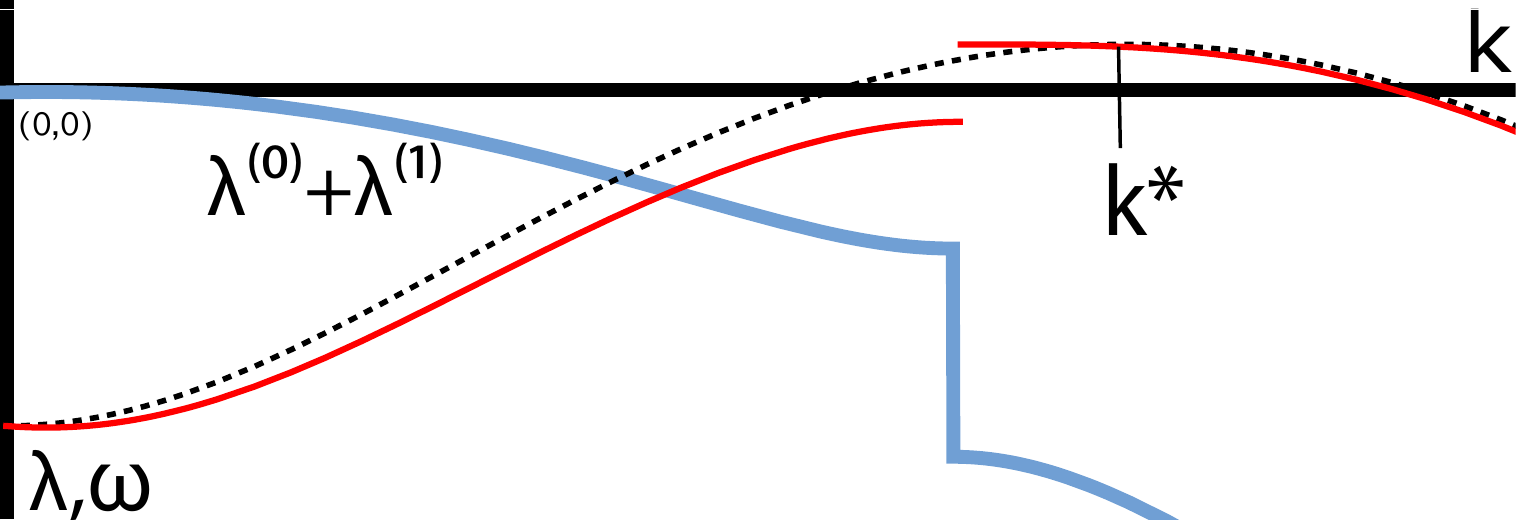}
  \vspace{5pt}
  \includegraphics[width=0.43\textwidth, height=0.3\textwidth]{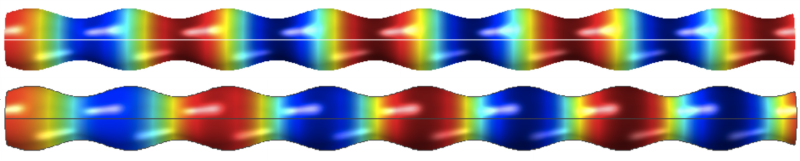}
  \includegraphics[width=0.042\textwidth]{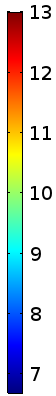}
  \caption{\label{fig:cylinderSpectrum} 
  (Top) three functions of wave-number: (blue) Laplacian eigenvalues
  on a rippled cylinder, including a band gap at the edge of the
  Brillouin zone (BZ) at $p/2$; (black dotted) Turing spectrum on a
  non-rippled cylinder; (red) Turing spectrum on a rippled cylinder.
  For the case shown with $k^*>p/2$, the sine mode is selected;
  (cosine selected if $k^*<p/2$). (Bottom) Numerical confirmation:
  Concentration patterns in the Thomas-Murray model switch from
  troughs to ridges as the unstable chemical wavelength, $k^*$, is
  dialed past twice the ripple wavelength by changing $\gamma = 1.125$ 
  (upper) to $\gamma = 0.975$ (lower).  Vertical dimension
  magnified 3x. Supplement includes video of sweeping $p$
  (Mathematica \& COMSOL files).
}
\end{figure}

This conformal mapping approach enables studies of other geometries, 
including bumps on spheres (below), drums (supplement), 
and rippled cylinders.
An axially symmetric rippled cylinder,
$h(z) = \cos\left[ (2\pi p/L) z \right] \equiv \cos(\bar{p} z)$, 
gives a Schr\"odinger-like equation at $\order{\epsilon}$ (cf. Eq.~\ref{geneigenvalue}):
\begin{equation}
  \begin{aligned}
    \left[-\Delta^0  - 2 \epsilon \lambda_{sk}^{(0)} \cos(\bar{p} v)\right] \phi_k = k^2 \phi_k \,.
    \label{eq:periodic}
   \end{aligned}
 \end{equation}
 This Schr{\"o}dinger equation describes a particle moving in a weak periodic
 potential, whose properties are well understood in the context of solid-state
 physics~\cite{kittel2004introduction}. At leading order, this perturbation 
 gives rise to a broken degeneracy (band-gap)
at $k = p/2$ with magnitude $4\epsilon \lambda_{sk}^{(0)}$.
Our analysis predicts that when an RD system governs surface concentrations,
the effective diffusion rate will increase in troughs and slow down on ridges.
Hence, diffusion is enhanced (diminished) 
where Gaussian curvature is negative (positive). 
This agrees with the short-time analysis of diffusion on Riemannian manifolds, 
where the leading order correction to diffusion is proportional to Gaussian curvature~\cite{David:1988,Faraudo:2002p5195, CastroVillarreal:2010p6177}. 
For Turing patterns, steady-state regions of
high concentration switch sharply from ridges to troughs as the most
unstable wavelength is dialed past twice the ripple wavelength, see Fig.~\ref{fig:cylinderSpectrum}.
Note that Eq.~\ref{periodic} predicts a broken degeneracy only when the most unstable wavelength is commensurate
with the ripple wavelength. However, in numerical simulations (presumably due to higher-order effects), we observe
pinning for a range of wavelengths close to commensurability, although the patterns become unpinned for sufficiently
incommensurate wavelengths.

An axially-symmetric distorted sphere, described by $R(\theta)$, has line element
\begin{equation}
  \dd{s}^2 = \left(R^2 + R'^2 \right) \dd{\theta}^2 + R^2 \sin^2\theta \dd{\varphi}^2\,,
  \nonumber
\end{equation}
where prime denotes derivatives.
Mapping to conformally flat coordinates:
\begin{equation}
  \dd{s}^2 = \Omega^2 \left(\dd{\Theta}^2 + \sin^2\Theta \dd{\varphi}^2 \right)\,.
  \nonumber
\end{equation}
The Laplacian eigenvalue equation in conformal coordinates becomes (see supplement)
\begin{equation}
  \left[-\Delta^0 +k^2 \left(R_0^2 - R^2 \frac{\sin^2 \theta}{\sin^2 \Theta} \right)\right] \Phi = k^2 R_0^2 \Phi\,,
  \label{spherelaplacian}
\end{equation}
where $\Delta^0$ is the Laplacian on a round sphere.
Setting $R = R_0\left(1 + \epsilon h(\theta)\right)$, we expand in powers of $\epsilon$.
To $\order{\epsilon}$,  $\theta = \Theta$, and Eq.~\eqref{spherelaplacian} reduces to
\begin{equation}\label{eq:CLsphere}
  \left[-\Delta^0 - 2 \epsilon k^2 R_0^2 h \right] \Phi = k^2 \Phi\,.
\end{equation}

\begin{figure}[h]
 \includegraphics[width=0.45\textwidth]{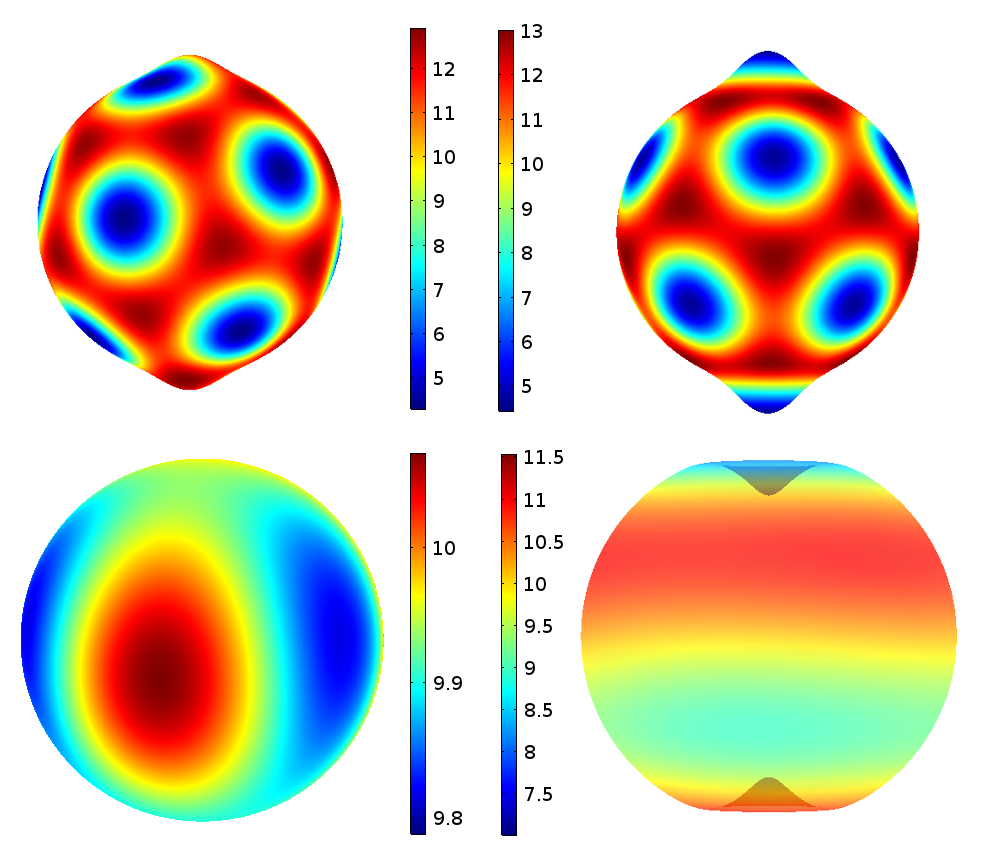}
  \caption{Sufficiently large bumps entrain spotted patterns (top pair).
  Initial patterning on an undeformed sphere shows $Y_5^5$ (bottom-left);
  the inward bumps of a ``pinched sphere'' 
  amplify lower harmonics, causing $Y_3^0$ to appear (bottom-right).
  \label{fig:sphere-bump}}
\end{figure}

Using Equation~\eqref{eq:CLsphere}, we study how deformations modify
the Laplacian eigenfunctions $Y_{\ell}^{m}(\theta,\phi)$. 
The perturbation depends only on $\theta$, breaking rotational symmetry by fixing polar
orientation of eigenmodes while preserving azimuthal symmetry. 
This can potentially entrain Turing patterns, although there are
competing influences from nonlinear effects and incommensurability of length
scales as seen in Fig.~\ref{fig:sphere-bump}. 
In fact, in numerical simulations, we often found that 
the initial patterning predicted by linear stability analysis 
would stabilize to more complex patterns.
In contrast, on a cylinder, initial patterning tends to persist indefinitely.
We attribute this to differing degrees of degeneracy. 
The spherical eigenmodes, $Y_{\ell}^{m}$, have  $2\ell+1$ degeneracy.
Thus, near the most unstable eigenmode, there are $2\ell$ additional unstable modes that can contribute to
non-linear pattern formation. 
In contrast, a cylinder presents only a twofold degeneracy.

As evident from Eq.~\ref{eq:CLsphere}, eigenfunction modifications are sensitive to the sign of $h$. 
An inward (outward) bump causes an increase (decrease) in diffusion. 
Non-linear effects in the Thomas-Murray model 
are known to stabilize spotted patterns instead of the single most unstable spherical harmonic, see Ref.~\cite{Varea:1999ur}. 
However, during the initial development of Turing patterns, 
the most unstable $Y_{\ell}^{m}$ is visible, and
the mode selected by a deformed sphere differs from that of an uniform sphere,
see Fig.~\ref{fig:sphere-bump}. 


\begin{figure}
  \includegraphics[width=0.45\textwidth]{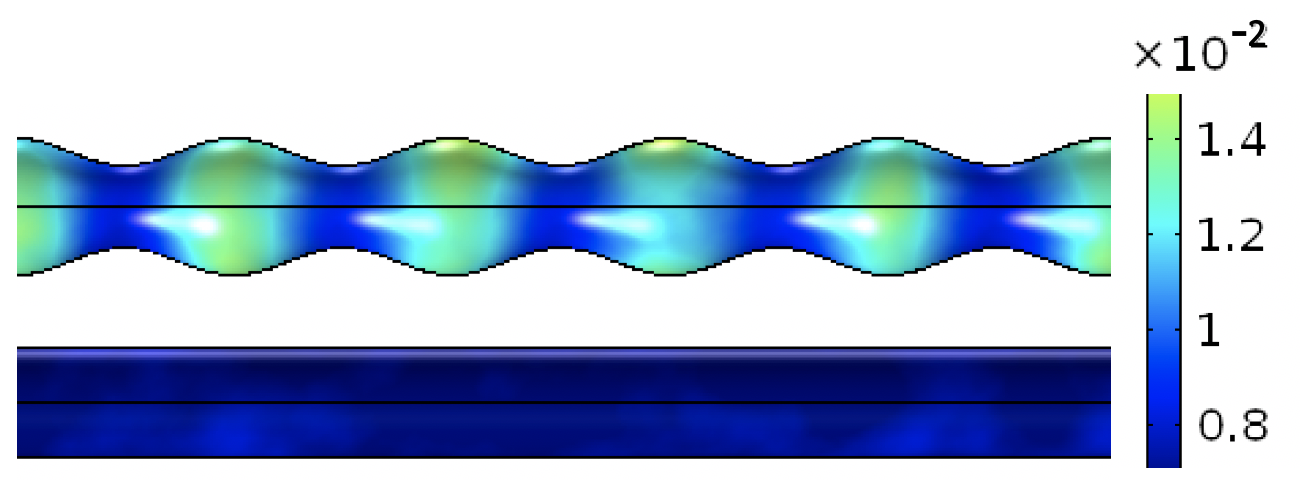}
  \caption{
	Noise-induced quasipatterns appear when geometric ripples split degenerate eigenmodes
	of an RD system below threshold.
    Deformations pin the time-averaged intensity of fluctuations in RD quasipatterns (top); 
	compare with the same intensity of extrinsic noise on a flat surface (bottom).
	For all two figures, $\nu_2 = 6$ and $K = 0.15$.}
  \label{fig:noisePatterns}
\end{figure}

Including extrinsic noise in RD systems leads to temporally fluctuating 
{\it quasi-patterns} for a broader range of parameters than those required for a Turing instability
~\cite{Lugo:2008, Butler:2011gh, Schumacher:2013, Biancalani:2017}. 
We have investigated how curvature influences such {\it noise-induced transient} patterns.
We add uncorrelated white noise $\eta(\vb{x}, t)$ 
of zero mean and variance $\mathcal{D}$ to Eq.~\ref{eqn:RD}.
In the linearly stable regime, this noise leads to eigenfunction fluctuations with a power spectrum
$P(\lambda)$ proportional to $\mathcal{D}\left[\text{det}\vb{M}(\lambda)\right]^{-1 }$.
On a uniform surface, time averaged fluctuations are translationally invariant,
$\langle \abs{\vb{\Psi}(x)}^2 \rangle = \vb{\Psi}_0^2$.
Deformations break this symmetry, leading to average fluctuations on a cylinder
of the form
\begin{equation}
  \langle \abs{\vb{\Psi}(x)}^2 \rangle \propto \sum_{k} \frac{\cos^{2}(\bar{k}x)}{\det\vb{M}(\lambda^{+}_{0k})}
  + \frac{\sin^{2}(\bar{k}x)}{\text{det}\vb{M}(\lambda^{-}_{0k})} + \order{\epsilon^2} \,.
  \label{eq:positionFluctuations}
\end{equation}
On spheres, the spherical harmonics replace the Fourier modes.
Figure~\ref{fig:noisePatterns} shows numerical verification that
such deformations fix the phase of fluctuations and create
non-uniform time-averaged intensities. 
Such behavior is expected to hold generically for
quasi-patterns induced by intrinsic noise~\cite{Lugo:2008, Butler:2011gh, Schumacher:2013, Biancalani:2017},
where the power spectrum describing fluctuations is qualitatively similar.

If these effects can be realized in manufactured or experimental systems 
they could enable systematic manipulation of patterns.
For example, collagen vitrigel (CV) for corneal endothelial regenerative treatment~\cite{yoshida2014development}
could be molded in a hemispherical or eggcrate geometry and used to support 
zebrafish chromatophores, which were recently cultured for \emph{in vitro} studies~\cite{yamanaka2014vitro} 
to explore RD models~\cite{nakamasu2009interactions}.
Mice hair follicle patterning is also driven by an RD system involving WNT growth factor~\cite{sick2006wnt}, and might be cultured on molded CV using established \emph{in vitro} techniques~\cite{philpott1990human}. 
Zebrafish stripes and hair follicles have spacing on the order of hundred microns.
Three orders of magnitude smaller,
cytoskeletal suspensions exhibit patterns~\cite{bois2011pattern, giomi2012polar, pearce2018geometrical}
that may realize our results.
For example, a pinched sphere might be constructed in the recently studied
encapsulations of actomyosin in giant unilamellar vesicles~\cite{durre2018capping}.
A recent review of such active fluids suggests that coupling to RD systems looks like a fruitful direction for research~\cite{marchetti2013hydrodynamics}.




Even in a system as simple as a cylinder with a ridge, geometry can
dramatically affect pattern formation. 
Our approach to the Laplacian on curved surfaces opens a new route to analytical understanding of patterns in real systems,
taking advantage of intuition and tools from quantum mechanics.
Future directions include pattern formation in the presence of advection~\cite{kumar2014pulsatory},
and on time-varying shapes.
Our approach can also be applied to scenarios involving interactions between
the surface and boundary physics. In particular, it would be interesting to examine 
the effect of nonuniform surface curvature when additional reaction-diffusion processes
take place at the boundary~\cite{Halatek2012}.
Our mathematical methods apply 
to any process described by the Laplacian, and
may also find application in 
soap films~\cite{osserman1986survey}, 
or Marangoni flows from surface tension gradients~\cite{Pereira:2007p5114}.

JRF acknowledges support from the Hertz Foundation.  JG acknowledges
support from CONACyT grant 180901.  MK
acknowledges support from DMR-1708280.

\bibliographystyle{apsrev4-1}
\bibliography{turing}

\begin{thebibliography}{47}%
\makeatletter
\providecommand \@ifxundefined [1]{%
 \@ifx{#1\undefined}
}%
\providecommand \@ifnum [1]{%
 \ifnum #1\expandafter \@firstoftwo
 \else \expandafter \@secondoftwo
 \fi
}%
\providecommand \@ifx [1]{%
 \ifx #1\expandafter \@firstoftwo
 \else \expandafter \@secondoftwo
 \fi
}%
\providecommand \natexlab [1]{#1}%
\providecommand \enquote  [1]{``#1''}%
\providecommand \bibnamefont  [1]{#1}%
\providecommand \bibfnamefont [1]{#1}%
\providecommand \citenamefont [1]{#1}%
\providecommand \href@noop [0]{\@secondoftwo}%
\providecommand \href [0]{\begingroup \@sanitize@url \@href}%
\providecommand \@href[1]{\@@startlink{#1}\@@href}%
\providecommand \@@href[1]{\endgroup#1\@@endlink}%
\providecommand \@sanitize@url [0]{\catcode `\\12\catcode `\$12\catcode
  `\&12\catcode `\#12\catcode `\^12\catcode `\_12\catcode `\%12\relax}%
\providecommand \@@startlink[1]{}%
\providecommand \@@endlink[0]{}%
\providecommand \url  [0]{\begingroup\@sanitize@url \@url }%
\providecommand \@url [1]{\endgroup\@href {#1}{\urlprefix }}%
\providecommand \urlprefix  [0]{URL }%
\providecommand \Eprint [0]{\href }%
\providecommand \doibase [0]{http://dx.doi.org/}%
\providecommand \selectlanguage [0]{\@gobble}%
\providecommand \bibinfo  [0]{\@secondoftwo}%
\providecommand \bibfield  [0]{\@secondoftwo}%
\providecommand \translation [1]{[#1]}%
\providecommand \BibitemOpen [0]{}%
\providecommand \bibitemStop [0]{}%
\providecommand \bibitemNoStop [0]{.\EOS\space}%
\providecommand \EOS [0]{\spacefactor3000\relax}%
\providecommand \BibitemShut  [1]{\csname bibitem#1\endcsname}%
\let\auto@bib@innerbib\@empty
\bibitem [{\citenamefont {Turing}(1952)}]{Turing:1952p2540}%
  \BibitemOpen
  \bibfield  {author} {\bibinfo {author} {\bibfnamefont {A.~M.}\ \bibnamefont
  {Turing}},\ }\href {http://www.jstor.org/stable/92463} {\bibfield  {journal}
  {\bibinfo  {journal} {Philosophical Transactions of the Royal Society of
  London. Series B, Biological Sciences}\ }\textbf {\bibinfo {volume} {237}},\
  \bibinfo {pages} {37} (\bibinfo {year} {1952})}\BibitemShut {NoStop}%
\bibitem [{\citenamefont {Hecht}\ \emph {et~al.}(2010)\citenamefont {Hecht},
  \citenamefont {Kessler},\ and\ \citenamefont {Levine}}]{hecht2010transient}%
  \BibitemOpen
  \bibfield  {author} {\bibinfo {author} {\bibfnamefont {I.}~\bibnamefont
  {Hecht}}, \bibinfo {author} {\bibfnamefont {D.~A.}\ \bibnamefont {Kessler}},
  \ and\ \bibinfo {author} {\bibfnamefont {H.}~\bibnamefont {Levine}},\ }\href
  {https://doi.org/10.1103/PhysRevLett.104.158301} {\bibfield  {journal}
  {\bibinfo  {journal} {Phys. Rev. Lett.}\ }\textbf {\bibinfo {volume} {104}},\
  \bibinfo {pages} {158301} (\bibinfo {year} {2010})}\BibitemShut {NoStop}%
\bibitem [{\citenamefont {Bois}\ \emph {et~al.}(2011)\citenamefont {Bois},
  \citenamefont {J{\"u}licher},\ and\ \citenamefont {Grill}}]{bois2011pattern}%
  \BibitemOpen
  \bibfield  {author} {\bibinfo {author} {\bibfnamefont {J.~S.}\ \bibnamefont
  {Bois}}, \bibinfo {author} {\bibfnamefont {F.}~\bibnamefont {J{\"u}licher}},
  \ and\ \bibinfo {author} {\bibfnamefont {S.~W.}\ \bibnamefont {Grill}},\
  }\href {https://doi.org/10.1103/PhysRevLett.106.028103} {\bibfield  {journal}
  {\bibinfo  {journal} {Phys. Rev. Lett.}\ }\textbf {\bibinfo {volume} {106}},\
  \bibinfo {pages} {028103} (\bibinfo {year} {2011})}\BibitemShut {NoStop}%
\bibitem [{\citenamefont {Giomi}\ and\ \citenamefont
  {Marchetti}(2012)}]{giomi2012polar}%
  \BibitemOpen
  \bibfield  {author} {\bibinfo {author} {\bibfnamefont {L.}~\bibnamefont
  {Giomi}}\ and\ \bibinfo {author} {\bibfnamefont {M.~C.}\ \bibnamefont
  {Marchetti}},\ }\href {\doibase 10.1039/c1sm06077e} {\bibfield  {journal}
  {\bibinfo  {journal} {Soft Matter}\ }\textbf {\bibinfo {volume} {8}},\
  \bibinfo {pages} {129} (\bibinfo {year} {2012})}\BibitemShut {NoStop}%
\bibitem [{\citenamefont {Pearce}\ \emph {et~al.}(2018)\citenamefont {Pearce},
  \citenamefont {Ellis}, \citenamefont {Fernandez-Nieves},\ and\ \citenamefont
  {Giomi}}]{pearce2018geometrical}%
  \BibitemOpen
  \bibfield  {author} {\bibinfo {author} {\bibfnamefont {D.}~\bibnamefont
  {Pearce}}, \bibinfo {author} {\bibfnamefont {P.~W.}\ \bibnamefont {Ellis}},
  \bibinfo {author} {\bibfnamefont {A.}~\bibnamefont {Fernandez-Nieves}}, \
  and\ \bibinfo {author} {\bibfnamefont {L.}~\bibnamefont {Giomi}},\ }\href
  {https://arxiv.org/abs/1805.01455} {\bibfield  {journal} {\bibinfo  {journal}
  {arXiv preprint arXiv:1805.01455}\ } (\bibinfo {year} {2018})}\BibitemShut
  {NoStop}%
\bibitem [{\citenamefont {D{\"u}rre}\ \emph {et~al.}(2018)\citenamefont
  {D{\"u}rre}, \citenamefont {Keber}, \citenamefont {Bleicher}, \citenamefont
  {Brauns}, \citenamefont {Cyron}, \citenamefont {Faix},\ and\ \citenamefont
  {Bausch}}]{durre2018capping}%
  \BibitemOpen
  \bibfield  {author} {\bibinfo {author} {\bibfnamefont {K.}~\bibnamefont
  {D{\"u}rre}}, \bibinfo {author} {\bibfnamefont {F.~C.}\ \bibnamefont
  {Keber}}, \bibinfo {author} {\bibfnamefont {P.}~\bibnamefont {Bleicher}},
  \bibinfo {author} {\bibfnamefont {F.}~\bibnamefont {Brauns}}, \bibinfo
  {author} {\bibfnamefont {C.~J.}\ \bibnamefont {Cyron}}, \bibinfo {author}
  {\bibfnamefont {J.}~\bibnamefont {Faix}}, \ and\ \bibinfo {author}
  {\bibfnamefont {A.~R.}\ \bibnamefont {Bausch}},\ }\href {\doibase
  10.1038/s41467-018-03918-1} {\bibfield  {journal} {\bibinfo  {journal}
  {Nature communications}\ }\textbf {\bibinfo {volume} {9}} (\bibinfo {year}
  {2018}),\ 10.1038/s41467-018-03918-1}\BibitemShut {NoStop}%
\bibitem [{\citenamefont {Marchetti}\ \emph {et~al.}(2013)\citenamefont
  {Marchetti}, \citenamefont {Joanny}, \citenamefont {Ramaswamy}, \citenamefont
  {Liverpool}, \citenamefont {Prost}, \citenamefont {Rao},\ and\ \citenamefont
  {Simha}}]{marchetti2013hydrodynamics}%
  \BibitemOpen
  \bibfield  {author} {\bibinfo {author} {\bibfnamefont {M.~C.}\ \bibnamefont
  {Marchetti}}, \bibinfo {author} {\bibfnamefont {J.-F.}\ \bibnamefont
  {Joanny}}, \bibinfo {author} {\bibfnamefont {S.}~\bibnamefont {Ramaswamy}},
  \bibinfo {author} {\bibfnamefont {T.~B.}\ \bibnamefont {Liverpool}}, \bibinfo
  {author} {\bibfnamefont {J.}~\bibnamefont {Prost}}, \bibinfo {author}
  {\bibfnamefont {M.}~\bibnamefont {Rao}}, \ and\ \bibinfo {author}
  {\bibfnamefont {R.~A.}\ \bibnamefont {Simha}},\ }\href {\doibase
  10.1103/RevModPhys.85.1143} {\bibfield  {journal} {\bibinfo  {journal}
  {Reviews of Modern Physics}\ }\textbf {\bibinfo {volume} {85}},\ \bibinfo
  {pages} {1143} (\bibinfo {year} {2013})}\BibitemShut {NoStop}%
\bibitem [{\citenamefont {Kumar}\ \emph {et~al.}(2014)\citenamefont {Kumar},
  \citenamefont {Bois}, \citenamefont {J{\"u}licher},\ and\ \citenamefont
  {Grill}}]{kumar2014pulsatory}%
  \BibitemOpen
  \bibfield  {author} {\bibinfo {author} {\bibfnamefont {K.~V.}\ \bibnamefont
  {Kumar}}, \bibinfo {author} {\bibfnamefont {J.~S.}\ \bibnamefont {Bois}},
  \bibinfo {author} {\bibfnamefont {F.}~\bibnamefont {J{\"u}licher}}, \ and\
  \bibinfo {author} {\bibfnamefont {S.~W.}\ \bibnamefont {Grill}},\ }\href
  {https://doi.org/10.1103/PhysRevLett.112.208101} {\bibfield  {journal}
  {\bibinfo  {journal} {Physical Review Letters}\ }\textbf {\bibinfo {volume}
  {112}},\ \bibinfo {pages} {208101} (\bibinfo {year} {2014})}\BibitemShut
  {NoStop}%
\bibitem [{\citenamefont {Almarcha}\ \emph {et~al.}(2010)\citenamefont
  {Almarcha}, \citenamefont {Trevelyan}, \citenamefont {Grosfils},\ and\
  \citenamefont {De~Wit}}]{almarcha2010chemically}%
  \BibitemOpen
  \bibfield  {author} {\bibinfo {author} {\bibfnamefont {C.}~\bibnamefont
  {Almarcha}}, \bibinfo {author} {\bibfnamefont {P.~M.}\ \bibnamefont
  {Trevelyan}}, \bibinfo {author} {\bibfnamefont {P.}~\bibnamefont {Grosfils}},
  \ and\ \bibinfo {author} {\bibfnamefont {A.}~\bibnamefont {De~Wit}},\ }\href
  {https://doi.org/10.1103/PhysRevLett.104.044501} {\bibfield  {journal}
  {\bibinfo  {journal} {Phys. Rev. Lett.}\ }\textbf {\bibinfo {volume} {104}},\
  \bibinfo {pages} {044501} (\bibinfo {year} {2010})}\BibitemShut {NoStop}%
\bibitem [{\citenamefont {Murray}(1981)}]{Murray:1981b}%
  \BibitemOpen
  \bibfield  {author} {\bibinfo {author} {\bibfnamefont {J.}~\bibnamefont
  {Murray}},\ }\href {\doibase https://doi.org/10.1016/0022-5193(81)90334-9}
  {\bibfield  {journal} {\bibinfo  {journal} {Journal of Theoretical Biology}\
  }\textbf {\bibinfo {volume} {88}},\ \bibinfo {pages} {161 } (\bibinfo {year}
  {1981})}\BibitemShut {NoStop}%
\bibitem [{\citenamefont {Butler}\ and\ \citenamefont
  {Goldenfeld}(2011)}]{Butler:2011gh}%
  \BibitemOpen
  \bibfield  {author} {\bibinfo {author} {\bibfnamefont {T.}~\bibnamefont
  {Butler}}\ and\ \bibinfo {author} {\bibfnamefont {N.}~\bibnamefont
  {Goldenfeld}},\ }\href {\doibase 10.1103/PhysRevE.84.011112} {\bibfield
  {journal} {\bibinfo  {journal} {Phys. Rev. E}\ }\textbf {\bibinfo {volume}
  {84}},\ \bibinfo {pages} {011112} (\bibinfo {year} {2011})}\BibitemShut
  {NoStop}%
\bibitem [{\citenamefont {Kondo}\ and\ \citenamefont
  {Miura}(2010)}]{kondo2010reaction}%
  \BibitemOpen
  \bibfield  {author} {\bibinfo {author} {\bibfnamefont {S.}~\bibnamefont
  {Kondo}}\ and\ \bibinfo {author} {\bibfnamefont {T.}~\bibnamefont {Miura}},\
  }\href {http://science.sciencemag.org/content/329/5999/1616} {\bibfield
  {journal} {\bibinfo  {journal} {science}\ }\textbf {\bibinfo {volume}
  {329}},\ \bibinfo {pages} {1616} (\bibinfo {year} {2010})}\BibitemShut
  {NoStop}%
\bibitem [{\citenamefont {Nakamasu}\ \emph {et~al.}(2009)\citenamefont
  {Nakamasu}, \citenamefont {Takahashi}, \citenamefont {Kanbe},\ and\
  \citenamefont {Kondo}}]{nakamasu2009interactions}%
  \BibitemOpen
  \bibfield  {author} {\bibinfo {author} {\bibfnamefont {A.}~\bibnamefont
  {Nakamasu}}, \bibinfo {author} {\bibfnamefont {G.}~\bibnamefont {Takahashi}},
  \bibinfo {author} {\bibfnamefont {A.}~\bibnamefont {Kanbe}}, \ and\ \bibinfo
  {author} {\bibfnamefont {S.}~\bibnamefont {Kondo}},\ }\href
  {http://www.pnas.org/content/106/21/8429} {\bibfield  {journal} {\bibinfo
  {journal} {Proceedings of the National Academy of Sciences}\ }\textbf
  {\bibinfo {volume} {106}},\ \bibinfo {pages} {8429} (\bibinfo {year}
  {2009})}\BibitemShut {NoStop}%
\bibitem [{\citenamefont {Yamanaka}\ and\ \citenamefont
  {Kondo}(2014)}]{yamanaka2014vitro}%
  \BibitemOpen
  \bibfield  {author} {\bibinfo {author} {\bibfnamefont {H.}~\bibnamefont
  {Yamanaka}}\ and\ \bibinfo {author} {\bibfnamefont {S.}~\bibnamefont
  {Kondo}},\ }\href {\doibase 10.1073/pnas.1315416111} {\bibfield  {journal}
  {\bibinfo  {journal} {Proceedings of the National Academy of Sciences}\
  }\textbf {\bibinfo {volume} {111}},\ \bibinfo {pages} {1867} (\bibinfo {year}
  {2014})}\BibitemShut {NoStop}%
\bibitem [{\citenamefont {Sasai}(2013)}]{sasai2013cytosystems}%
  \BibitemOpen
  \bibfield  {author} {\bibinfo {author} {\bibfnamefont {Y.}~\bibnamefont
  {Sasai}},\ }\href {https://www.nature.com/articles/nature11859} {\bibfield
  {journal} {\bibinfo  {journal} {Nature}\ }\textbf {\bibinfo {volume} {493}},\
  \bibinfo {pages} {318} (\bibinfo {year} {2013})}\BibitemShut {NoStop}%
\bibitem [{\citenamefont {Biancalani}\ \emph {et~al.}(2017)\citenamefont
  {Biancalani}, \citenamefont {Jafarpour},\ and\ \citenamefont
  {Goldenfeld}}]{Biancalani:2017}%
  \BibitemOpen
  \bibfield  {author} {\bibinfo {author} {\bibfnamefont {T.}~\bibnamefont
  {Biancalani}}, \bibinfo {author} {\bibfnamefont {F.}~\bibnamefont
  {Jafarpour}}, \ and\ \bibinfo {author} {\bibfnamefont {N.}~\bibnamefont
  {Goldenfeld}},\ }\href {\doibase 10.1103/PhysRevLett.118.018101} {\bibfield
  {journal} {\bibinfo  {journal} {Phys. Rev. Lett.}\ }\textbf {\bibinfo
  {volume} {118}},\ \bibinfo {pages} {018101} (\bibinfo {year}
  {2017})}\BibitemShut {NoStop}%
\bibitem [{\citenamefont {Corson}\ \emph {et~al.}(2017)\citenamefont {Corson},
  \citenamefont {Couturier}, \citenamefont {Rouault}, \citenamefont {Mazouni},\
  and\ \citenamefont {Schweisguth}}]{Corson:2017}%
  \BibitemOpen
  \bibfield  {author} {\bibinfo {author} {\bibfnamefont {F.}~\bibnamefont
  {Corson}}, \bibinfo {author} {\bibfnamefont {L.}~\bibnamefont {Couturier}},
  \bibinfo {author} {\bibfnamefont {H.}~\bibnamefont {Rouault}}, \bibinfo
  {author} {\bibfnamefont {K.}~\bibnamefont {Mazouni}}, \ and\ \bibinfo
  {author} {\bibfnamefont {F.}~\bibnamefont {Schweisguth}},\ }\href
  {https://science.sciencemag.org/content/356/6337/eaai7407} {\bibfield
  {journal} {\bibinfo  {journal} {Science}\ }\textbf {\bibinfo {volume} {356}}
  (\bibinfo {year} {2017})}\BibitemShut {NoStop}%
\bibitem [{\citenamefont {Isaac}\ \emph {et~al.}(2013)\citenamefont {Isaac},
  \citenamefont {Manor}, \citenamefont {Kachar}, \citenamefont {Yochelis},\
  and\ \citenamefont {Gov}}]{isaac2013linking}%
  \BibitemOpen
  \bibfield  {author} {\bibinfo {author} {\bibfnamefont {E.~B.}\ \bibnamefont
  {Isaac}}, \bibinfo {author} {\bibfnamefont {U.}~\bibnamefont {Manor}},
  \bibinfo {author} {\bibfnamefont {B.}~\bibnamefont {Kachar}}, \bibinfo
  {author} {\bibfnamefont {A.}~\bibnamefont {Yochelis}}, \ and\ \bibinfo
  {author} {\bibfnamefont {N.~S.}\ \bibnamefont {Gov}},\ }\href
  {https://doi.org/10.1103/PhysRevE.88.022718} {\bibfield  {journal} {\bibinfo
  {journal} {Phys. Rev. E}\ }\textbf {\bibinfo {volume} {88}},\ \bibinfo
  {pages} {022718} (\bibinfo {year} {2013})}\BibitemShut {NoStop}%
\bibitem [{\citenamefont {Turner}\ \emph {et~al.}(2010)\citenamefont {Turner},
  \citenamefont {Vitelli},\ and\ \citenamefont {Nelson}}]{turner2010vortices}%
  \BibitemOpen
  \bibfield  {author} {\bibinfo {author} {\bibfnamefont {A.~M.}\ \bibnamefont
  {Turner}}, \bibinfo {author} {\bibfnamefont {V.}~\bibnamefont {Vitelli}}, \
  and\ \bibinfo {author} {\bibfnamefont {D.~R.}\ \bibnamefont {Nelson}},\
  }\href {https://doi.org/10.1103/RevModPhys.82.1301} {\bibfield  {journal}
  {\bibinfo  {journal} {Reviews of Modern Physics}\ }\textbf {\bibinfo {volume}
  {82}},\ \bibinfo {pages} {1301} (\bibinfo {year} {2010})}\BibitemShut
  {NoStop}%
\bibitem [{\citenamefont {Fialho}\ \emph {et~al.}(2017)\citenamefont {Fialho},
  \citenamefont {Bernardino}, \citenamefont {Silvestre},\ and\ \citenamefont
  {Telo~da Gama}}]{Fialho:2017}%
  \BibitemOpen
  \bibfield  {author} {\bibinfo {author} {\bibfnamefont {A.~R.}\ \bibnamefont
  {Fialho}}, \bibinfo {author} {\bibfnamefont {N.~R.}\ \bibnamefont
  {Bernardino}}, \bibinfo {author} {\bibfnamefont {N.~M.}\ \bibnamefont
  {Silvestre}}, \ and\ \bibinfo {author} {\bibfnamefont {M.~M.}\ \bibnamefont
  {Telo~da Gama}},\ }\href {\doibase 10.1103/PhysRevE.95.012702} {\bibfield
  {journal} {\bibinfo  {journal} {Phys. Rev. E}\ }\textbf {\bibinfo {volume}
  {95}},\ \bibinfo {pages} {012702} (\bibinfo {year} {2017})}\BibitemShut
  {NoStop}%
\bibitem [{\citenamefont {Shankar}\ \emph {et~al.}(2017)\citenamefont
  {Shankar}, \citenamefont {Bowick},\ and\ \citenamefont
  {Marchetti}}]{Shankar:2017}%
  \BibitemOpen
  \bibfield  {author} {\bibinfo {author} {\bibfnamefont {S.}~\bibnamefont
  {Shankar}}, \bibinfo {author} {\bibfnamefont {M.~J.}\ \bibnamefont {Bowick}},
  \ and\ \bibinfo {author} {\bibfnamefont {M.~C.}\ \bibnamefont {Marchetti}},\
  }\href {\doibase 10.1103/PhysRevX.7.031039} {\bibfield  {journal} {\bibinfo
  {journal} {Phys. Rev. X}\ }\textbf {\bibinfo {volume} {7}},\ \bibinfo {pages}
  {031039} (\bibinfo {year} {2017})}\BibitemShut {NoStop}%
\bibitem [{\citenamefont {Miller}\ \emph {et~al.}(2018)\citenamefont {Miller},
  \citenamefont {Stoop},\ and\ \citenamefont
  {Dunkel}}]{PhysRevLett.120.268001}%
  \BibitemOpen
  \bibfield  {author} {\bibinfo {author} {\bibfnamefont {P.~W.}\ \bibnamefont
  {Miller}}, \bibinfo {author} {\bibfnamefont {N.}~\bibnamefont {Stoop}}, \
  and\ \bibinfo {author} {\bibfnamefont {J.}~\bibnamefont {Dunkel}},\ }\href
  {\doibase 10.1103/PhysRevLett.120.268001} {\bibfield  {journal} {\bibinfo
  {journal} {Phys. Rev. Lett.}\ }\textbf {\bibinfo {volume} {120}},\ \bibinfo
  {pages} {268001} (\bibinfo {year} {2018})}\BibitemShut {NoStop}%
\bibitem [{\citenamefont {Thalmeier}\ \emph {et~al.}(2016)\citenamefont
  {Thalmeier}, \citenamefont {Halatek},\ and\ \citenamefont
  {Frey}}]{Thalmeier:2016}%
  \BibitemOpen
  \bibfield  {author} {\bibinfo {author} {\bibfnamefont {D.}~\bibnamefont
  {Thalmeier}}, \bibinfo {author} {\bibfnamefont {J.}~\bibnamefont {Halatek}},
  \ and\ \bibinfo {author} {\bibfnamefont {E.}~\bibnamefont {Frey}},\ }\href
  {\doibase 10.1073/pnas.1515191113} {\bibfield  {journal} {\bibinfo  {journal}
  {Proceedings of the National Academy of Sciences}\ }\textbf {\bibinfo
  {volume} {113}},\ \bibinfo {pages} {548} (\bibinfo {year}
  {2016})}\BibitemShut {NoStop}%
\bibitem [{\citenamefont {Vandin}\ \emph {et~al.}(2016)\citenamefont {Vandin},
  \citenamefont {Marenduzzo}, \citenamefont {Goryachev},\ and\ \citenamefont
  {Orlandini}}]{Vandin:2016}%
  \BibitemOpen
  \bibfield  {author} {\bibinfo {author} {\bibfnamefont {G.}~\bibnamefont
  {Vandin}}, \bibinfo {author} {\bibfnamefont {D.}~\bibnamefont {Marenduzzo}},
  \bibinfo {author} {\bibfnamefont {A.~B.}\ \bibnamefont {Goryachev}}, \ and\
  \bibinfo {author} {\bibfnamefont {E.}~\bibnamefont {Orlandini}},\ }\href
  {\doibase 10.1039/C6SM00340K} {\bibfield  {journal} {\bibinfo  {journal}
  {Soft Matter}\ }\textbf {\bibinfo {volume} {12}},\ \bibinfo {pages} {3888}
  (\bibinfo {year} {2016})}\BibitemShut {NoStop}%
\bibitem [{\citenamefont {Yu}\ \emph {et~al.}(2017)\citenamefont {Yu},
  \citenamefont {Wang}, \citenamefont {Ni}, \citenamefont {He}, \citenamefont
  {Huang}, \citenamefont {Lin}, \citenamefont {Qian},\ and\ \citenamefont
  {Jiang}}]{Yu:2017}%
  \BibitemOpen
  \bibfield  {author} {\bibinfo {author} {\bibfnamefont {S.}~\bibnamefont
  {Yu}}, \bibinfo {author} {\bibfnamefont {H.}~\bibnamefont {Wang}}, \bibinfo
  {author} {\bibfnamefont {Y.}~\bibnamefont {Ni}}, \bibinfo {author}
  {\bibfnamefont {L.}~\bibnamefont {He}}, \bibinfo {author} {\bibfnamefont
  {M.}~\bibnamefont {Huang}}, \bibinfo {author} {\bibfnamefont
  {Y.}~\bibnamefont {Lin}}, \bibinfo {author} {\bibfnamefont {J.}~\bibnamefont
  {Qian}}, \ and\ \bibinfo {author} {\bibfnamefont {H.}~\bibnamefont {Jiang}},\
  }\href {\doibase 10.1039/C7SM01278K} {\bibfield  {journal} {\bibinfo
  {journal} {Soft Matter}\ }\textbf {\bibinfo {volume} {13}},\ \bibinfo {pages}
  {5970} (\bibinfo {year} {2017})}\BibitemShut {NoStop}%
\bibitem [{\citenamefont {Nampoothiri}(2016)}]{Nampoothiri:2016}%
  \BibitemOpen
  \bibfield  {author} {\bibinfo {author} {\bibfnamefont {S.}~\bibnamefont
  {Nampoothiri}},\ }\href {\doibase 10.1103/PhysRevE.94.022403} {\bibfield
  {journal} {\bibinfo  {journal} {Phys. Rev. E}\ }\textbf {\bibinfo {volume}
  {94}},\ \bibinfo {pages} {022403} (\bibinfo {year} {2016})}\BibitemShut
  {NoStop}%
\bibitem [{\citenamefont {S{\'a}nchez-Gardu{\~n}o}\ \emph
  {et~al.}(2018)\citenamefont {S{\'a}nchez-Gardu{\~n}o}, \citenamefont
  {Krause}, \citenamefont {Castillo},\ and\ \citenamefont
  {Padilla}}]{Sanchez-Garduno:2018}%
  \BibitemOpen
  \bibfield  {author} {\bibinfo {author} {\bibfnamefont {F.}~\bibnamefont
  {S{\'a}nchez-Gardu{\~n}o}}, \bibinfo {author} {\bibfnamefont {A.~L.}\
  \bibnamefont {Krause}}, \bibinfo {author} {\bibfnamefont {J.~A.}\
  \bibnamefont {Castillo}}, \ and\ \bibinfo {author} {\bibfnamefont
  {P.}~\bibnamefont {Padilla}},\ }\href
  {https://doi.org/10.1016/j.jtbi.2018.09.028} {\bibfield  {journal} {\bibinfo
  {journal} {Journal of Theoretical Biology}\ } (\bibinfo {year}
  {2018})}\BibitemShut {NoStop}%
\bibitem [{\citenamefont {{Nampoothiri}}\ and\ \citenamefont
  {{Medhi}}(2017)}]{Nampoothiri:2017}%
  \BibitemOpen
  \bibfield  {author} {\bibinfo {author} {\bibfnamefont {S.}~\bibnamefont
  {{Nampoothiri}}}\ and\ \bibinfo {author} {\bibfnamefont {A.}~\bibnamefont
  {{Medhi}}},\ }\href@noop {} {\bibfield  {journal} {\bibinfo  {journal} {ArXiv
  e-prints}\ } (\bibinfo {year} {2017})},\ \Eprint
  {http://arxiv.org/abs/1705.02119} {arXiv:1705.02119 [cond-mat.soft]}
  \BibitemShut {NoStop}%
\bibitem [{\citenamefont {Kac}(1966)}]{Kac:1966p4109}%
  \BibitemOpen
  \bibfield  {author} {\bibinfo {author} {\bibfnamefont {M.}~\bibnamefont
  {Kac}},\ }\href {http://www.jstor.org/stable/2313748} {\bibfield  {journal}
  {\bibinfo  {journal} {The American Mathematical Monthly}\ }\textbf {\bibinfo
  {volume} {73}},\ \bibinfo {pages} {1} (\bibinfo {year} {1966})}\BibitemShut
  {NoStop}%
\bibitem [{\citenamefont {Schoen}\ and\ \citenamefont
  {Yau}(1994)}]{Schoen:1994}%
  \BibitemOpen
  \bibfield  {author} {\bibinfo {author} {\bibfnamefont {R.}~\bibnamefont
  {Schoen}}\ and\ \bibinfo {author} {\bibfnamefont {S.}~\bibnamefont {Yau}},\
  }\href {https://books.google.com/books?id=d4VtQgAACAAJ} {\emph {\bibinfo
  {title} {Lectures on Differential Geometry}}},\ Conference Proceedings and
  Lecture Note\ (\bibinfo  {publisher} {International Press},\ \bibinfo {year}
  {1994})\BibitemShut {NoStop}%
\bibitem [{\citenamefont {Rosenberg}(1997)}]{Rosenberg:1997}%
  \BibitemOpen
  \bibfield  {author} {\bibinfo {author} {\bibfnamefont {S.}~\bibnamefont
  {Rosenberg}},\ }\href {\doibase 10.1017/CBO9780511623783} {\emph {\bibinfo
  {title} {The Laplacian on a Riemannian Manifold: An Introduction to Analysis
  on Manifolds}}},\ London Mathematical Society Student Texts\ (\bibinfo
  {publisher} {Cambridge University Press},\ \bibinfo {year}
  {1997})\BibitemShut {NoStop}%
\bibitem [{\citenamefont {Braun}\ \emph {et~al.}(2008)\citenamefont {Braun},
  \citenamefont {Brelidze}, \citenamefont {Douglas},\ and\ \citenamefont
  {Ovrut}}]{Braun:2008p4810}%
  \BibitemOpen
  \bibfield  {author} {\bibinfo {author} {\bibfnamefont {V.}~\bibnamefont
  {Braun}}, \bibinfo {author} {\bibfnamefont {T.}~\bibnamefont {Brelidze}},
  \bibinfo {author} {\bibfnamefont {M.~R.}\ \bibnamefont {Douglas}}, \ and\
  \bibinfo {author} {\bibfnamefont {B.~A.}\ \bibnamefont {Ovrut}},\ }\href
  {http://stacks.iop.org/1126-6708/2008/i=05/a=080} {\bibfield  {journal}
  {\bibinfo  {journal} {Journal of High Energy Physics}\ }\textbf {\bibinfo
  {volume} {2008}},\ \bibinfo {pages} {080} (\bibinfo {year}
  {2008})}\BibitemShut {NoStop}%
\bibitem [{\citenamefont {David}(1987)}]{David:1988}%
  \BibitemOpen
  \bibfield  {author} {\bibinfo {author} {\bibfnamefont {F.}~\bibnamefont
  {David}},\ }in\ \href@noop {} {\emph {\bibinfo {booktitle} {{5th Jerusalem
  Winter School for Theoretical Physics}}}}\ (\bibinfo {year} {1987})\ pp.\
  \bibinfo {pages} {157--223}\BibitemShut {NoStop}%
\bibitem [{\citenamefont {Faraudo}(2002)}]{Faraudo:2002p5195}%
  \BibitemOpen
  \bibfield  {author} {\bibinfo {author} {\bibfnamefont {J.}~\bibnamefont
  {Faraudo}},\ }\href {\doibase 10.1063/1.1456024} {\bibfield  {journal}
  {\bibinfo  {journal} {The Journal of Chemical Physics}\ }\textbf {\bibinfo
  {volume} {116}},\ \bibinfo {pages} {5831} (\bibinfo {year}
  {2002})}\BibitemShut {NoStop}%
\bibitem [{\citenamefont
  {Castro-Villarreal}(2010)}]{CastroVillarreal:2010p6177}%
  \BibitemOpen
  \bibfield  {author} {\bibinfo {author} {\bibfnamefont {P.}~\bibnamefont
  {Castro-Villarreal}},\ }\href
  {http://stacks.iop.org/1742-5468/2010/i=08/a=P08006} {\bibfield  {journal}
  {\bibinfo  {journal} {Journal of Statistical Mechanics: Theory and
  Experiment}\ }\textbf {\bibinfo {volume} {2010}},\ \bibinfo {pages} {P08006}
  (\bibinfo {year} {2010})}\BibitemShut {NoStop}%
\bibitem [{\citenamefont {COMSOL}()}]{comsol}%
  \BibitemOpen
  \bibfield  {author} {\bibinfo {author} {\bibnamefont {COMSOL}},\ }\href
  {www.comsol.com} {\enquote {\bibinfo {title} {Multiphysics v5.2a},}\
  }\BibitemShut {NoStop}%
\bibitem [{\citenamefont {da~Costa}(1981)}]{daCosta:1982}%
  \BibitemOpen
  \bibfield  {author} {\bibinfo {author} {\bibfnamefont {R.~C.~T.}\
  \bibnamefont {da~Costa}},\ }\href {\doibase 10.1103/PhysRevA.23.1982}
  {\bibfield  {journal} {\bibinfo  {journal} {Phys. Rev. A}\ }\textbf {\bibinfo
  {volume} {23}},\ \bibinfo {pages} {1982} (\bibinfo {year}
  {1981})}\BibitemShut {NoStop}%
\bibitem [{\citenamefont {Kittel}(2004)}]{kittel2004introduction}%
  \BibitemOpen
  \bibfield  {author} {\bibinfo {author} {\bibfnamefont {C.}~\bibnamefont
  {Kittel}},\ }\href@noop {} {\emph {\bibinfo {title} {Introduction to Solid
  State Physics}}}\ (\bibinfo  {publisher} {John Wiley \& Sons},\ \bibinfo
  {year} {2004})\BibitemShut {NoStop}%
\bibitem [{\citenamefont {Barrio}\ \emph {et~al.}(1999)\citenamefont {Barrio},
  \citenamefont {Varea}, \citenamefont {Arag{\'o}n},\ and\ \citenamefont
  {Maini}}]{Varea:1999ur}%
  \BibitemOpen
  \bibfield  {author} {\bibinfo {author} {\bibfnamefont {R.~A.}\ \bibnamefont
  {Barrio}}, \bibinfo {author} {\bibfnamefont {C.}~\bibnamefont {Varea}},
  \bibinfo {author} {\bibfnamefont {J.~L.}\ \bibnamefont {Arag{\'o}n}}, \ and\
  \bibinfo {author} {\bibfnamefont {P.~K.}\ \bibnamefont {Maini}},\ }\href
  {\doibase 10.1006/bulm.1998.0093} {\bibfield  {journal} {\bibinfo  {journal}
  {Bulletin of Mathematical Biology}\ }\textbf {\bibinfo {volume} {61}},\
  \bibinfo {pages} {483} (\bibinfo {year} {1999})}\BibitemShut {NoStop}%
\bibitem [{\citenamefont {Lugo}\ and\ \citenamefont
  {McKane}(2008)}]{Lugo:2008}%
  \BibitemOpen
  \bibfield  {author} {\bibinfo {author} {\bibfnamefont {C.~A.}\ \bibnamefont
  {Lugo}}\ and\ \bibinfo {author} {\bibfnamefont {A.~J.}\ \bibnamefont
  {McKane}},\ }\href {\doibase 10.1103/PhysRevE.78.051911} {\bibfield
  {journal} {\bibinfo  {journal} {Phys. Rev. E}\ }\textbf {\bibinfo {volume}
  {78}},\ \bibinfo {pages} {051911} (\bibinfo {year} {2008})}\BibitemShut
  {NoStop}%
\bibitem [{\citenamefont {Schumacher}\ \emph {et~al.}(2013)\citenamefont
  {Schumacher}, \citenamefont {Woolley},\ and\ \citenamefont
  {Baker}}]{Schumacher:2013}%
  \BibitemOpen
  \bibfield  {author} {\bibinfo {author} {\bibfnamefont {L.~J.}\ \bibnamefont
  {Schumacher}}, \bibinfo {author} {\bibfnamefont {T.~E.}\ \bibnamefont
  {Woolley}}, \ and\ \bibinfo {author} {\bibfnamefont {R.~E.}\ \bibnamefont
  {Baker}},\ }\href {\doibase 10.1103/PhysRevE.87.042719} {\bibfield  {journal}
  {\bibinfo  {journal} {Phys. Rev. E}\ }\textbf {\bibinfo {volume} {87}},\
  \bibinfo {pages} {042719} (\bibinfo {year} {2013})}\BibitemShut {NoStop}%
\bibitem [{\citenamefont {Yoshida}\ \emph {et~al.}(2014)\citenamefont
  {Yoshida}, \citenamefont {Oshikata-Miyazaki}, \citenamefont {Yokoo},
  \citenamefont {Yamagami}, \citenamefont {Takezawa},\ and\ \citenamefont
  {Amano}}]{yoshida2014development}%
  \BibitemOpen
  \bibfield  {author} {\bibinfo {author} {\bibfnamefont {J.}~\bibnamefont
  {Yoshida}}, \bibinfo {author} {\bibfnamefont {A.}~\bibnamefont
  {Oshikata-Miyazaki}}, \bibinfo {author} {\bibfnamefont {S.}~\bibnamefont
  {Yokoo}}, \bibinfo {author} {\bibfnamefont {S.}~\bibnamefont {Yamagami}},
  \bibinfo {author} {\bibfnamefont {T.}~\bibnamefont {Takezawa}}, \ and\
  \bibinfo {author} {\bibfnamefont {S.}~\bibnamefont {Amano}},\ }\href
  {\doibase 10.1167/iovs.14-14211} {\bibfield  {journal} {\bibinfo  {journal}
  {Investigative ophthalmology \& visual science}\ }\textbf {\bibinfo {volume}
  {55}},\ \bibinfo {pages} {4975} (\bibinfo {year} {2014})}\BibitemShut
  {NoStop}%
\bibitem [{\citenamefont {Sick}\ \emph {et~al.}(2006)\citenamefont {Sick},
  \citenamefont {Reinker}, \citenamefont {Timmer},\ and\ \citenamefont
  {Schlake}}]{sick2006wnt}%
  \BibitemOpen
  \bibfield  {author} {\bibinfo {author} {\bibfnamefont {S.}~\bibnamefont
  {Sick}}, \bibinfo {author} {\bibfnamefont {S.}~\bibnamefont {Reinker}},
  \bibinfo {author} {\bibfnamefont {J.}~\bibnamefont {Timmer}}, \ and\ \bibinfo
  {author} {\bibfnamefont {T.}~\bibnamefont {Schlake}},\ }\href
  {http://science.sciencemag.org/content/314/5804/1447} {\bibfield  {journal}
  {\bibinfo  {journal} {Science}\ }\textbf {\bibinfo {volume} {314}},\ \bibinfo
  {pages} {1447} (\bibinfo {year} {2006})}\BibitemShut {NoStop}%
\bibitem [{\citenamefont {Philpott}\ \emph {et~al.}(1990)\citenamefont
  {Philpott}, \citenamefont {Green},\ and\ \citenamefont
  {Kealey}}]{philpott1990human}%
  \BibitemOpen
  \bibfield  {author} {\bibinfo {author} {\bibfnamefont {M.~P.}\ \bibnamefont
  {Philpott}}, \bibinfo {author} {\bibfnamefont {M.~R.}\ \bibnamefont {Green}},
  \ and\ \bibinfo {author} {\bibfnamefont {T.}~\bibnamefont {Kealey}},\ }\href
  {http://jcs.biologists.org/content/joces/97/3/463.full.pdf} {\bibfield
  {journal} {\bibinfo  {journal} {Journal of cell science}\ }\textbf {\bibinfo
  {volume} {97}},\ \bibinfo {pages} {463} (\bibinfo {year} {1990})}\BibitemShut
  {NoStop}%
\bibitem [{\citenamefont {Halatek}\ and\ \citenamefont
  {Frey}(2012)}]{Halatek2012}%
  \BibitemOpen
  \bibfield  {author} {\bibinfo {author} {\bibfnamefont {J.}~\bibnamefont
  {Halatek}}\ and\ \bibinfo {author} {\bibfnamefont {E.}~\bibnamefont {Frey}},\
  }\href {\doibase 10.1016/j.celrep.2012.04.005} {\bibfield  {journal}
  {\bibinfo  {journal} {Cell reports}\ }\textbf {\bibinfo {volume} {1}},\
  \bibinfo {pages} {741} (\bibinfo {year} {2012})}\BibitemShut {NoStop}%
\bibitem [{\citenamefont {Osserman}(2002)}]{osserman1986survey}%
  \BibitemOpen
  \bibfield  {author} {\bibinfo {author} {\bibfnamefont {R.}~\bibnamefont
  {Osserman}},\ }\href {https://books.google.com/books?id=WHN444vBvioC} {\emph
  {\bibinfo {title} {A Survey of Minimal Surfaces}}},\ Dover Phoenix editions\
  (\bibinfo  {publisher} {Dover Publications},\ \bibinfo {year}
  {2002})\BibitemShut {NoStop}%
\bibitem [{\citenamefont {Pereira}\ \emph {et~al.}(2007)\citenamefont
  {Pereira}, \citenamefont {Trevelyan}, \citenamefont {Thiele},\ and\
  \citenamefont {Kalliadasis}}]{Pereira:2007p5114}%
  \BibitemOpen
  \bibfield  {author} {\bibinfo {author} {\bibfnamefont {A.}~\bibnamefont
  {Pereira}}, \bibinfo {author} {\bibfnamefont {P.~M.~J.}\ \bibnamefont
  {Trevelyan}}, \bibinfo {author} {\bibfnamefont {U.}~\bibnamefont {Thiele}}, \
  and\ \bibinfo {author} {\bibfnamefont {S.}~\bibnamefont {Kalliadasis}},\
  }\href {\doibase 10.1063/1.2775938} {\bibfield  {journal} {\bibinfo
  {journal} {Physics of Fluids}\ }\textbf {\bibinfo {volume} {19}},\ \bibinfo
  {pages} {112102} (\bibinfo {year} {2007})}\BibitemShut {NoStop}%
\end{thebibliography}%

\end{document}